\documentclass[12pt]{iopart}
\usepackage{graphicx}% Include figure files
\usepackage{dcolumn}% Align table columns on decimal point
\usepackage{bm}% bold math
\usepackage{iopams} 

\usepackage{epsfig}

\begin{document}

\title{ Evolution equation for soft physics at high energy }%

\author{P. Brogueira}
\address{Departamento de F\' \i sica, IST, Av. Rovisco Pais, 1049-001 Lisboa, Portugal }
\ead{pedro@fisica.ist.utl.pt}
\author{J. Dias de Deus}
\address{CENTRA, Departamento de F\' \i sica, IST, Av. Rovisco Pais, 1049-001 Lisboa, Portugal}
\ead{jdd@fisica.ist.utl.pt}

\begin{abstract}
Based on the non-linear logistic equation we study, in a qualitative and semi-quantitative way, the evolution with energy and saturation of the elastic differential cross-section in $pp(\bar{p}p)$ collisions at high energy. Geometrical scaling occurs at the black disk limit, and scaling develops first for small values of the scaling variable $|t|\sigma_{tot.}$. Our prediction for $d \sigma / \ d t$ at LHC, with two zeros and a minimum at large $|t|$ differs, as far as we know, from all existing ones.
\end{abstract}

 % end of PACS codes
\maketitle

Saturation phenomena are expected  to dominate QCD physics at high energy and high matter density \cite{ref1,ref2}, as it may happen at LHC and cosmic rays at ultra high energies. Non linear differential equations include, in a natural way, saturation effects. This happens with the well known logistic equation\cite{ref3}, which can be seen as a simplified version of the B-K equation\cite{ref4}. See\cite{ref5} and \cite{ref6} for discussions on evolution and saturation.

We shall concentrate here in the evolution of the imaginary part of the impact parameter elastic amplitude, or the profile function $\Gamma (b,R) \equiv {\rm Im} \, B(b,R)$, where $b$ is the impact parameter, related to angular momentum $\ell$ by
\begin{equation}
\label{equa1}
b \simeq \frac{2}{\sqrt{s}}\ell,
\end{equation}
where $\sqrt{s}$ is the centre of mass energy, and $R$ is an increasing with energy radial scale parameter. Partial wave unitarity constrains $\Gamma (b,R)$:
\begin{equation}
\label{equa2}
0 \leq \Gamma (b,R) \leq 1.
\end{equation}

We now write two logistic equations, in $R$ and $b$, respectively:
\begin{equation}
\label{equa3}
\frac{\partial \Gamma }{\partial R}= \frac{1}{\gamma}(\Gamma -\Gamma ^2),
\end{equation}
and
\begin{equation}
\label{equa4}
\frac{\partial \Gamma }{\partial b}= -\frac{1}{\gamma}(\Gamma -\Gamma ^2),
\end{equation}
where $\gamma > 0$ is pratically a constant. From \eref{equa3} one sees that  $\partial \Gamma  / \partial R >0$ and that, for fixed $b$ and $\Gamma >0$, $\Gamma $ reaches the black disk limit: $\Gamma =\Gamma ^2=1$. From\eref{equa4} one sees that in general $\Gamma $ is a decreasing function of $b$ and that, for large $b$, $\Gamma $ decreases, as expected, exponentially $(\Gamma  \sim \exp{-b/ \gamma})$, saturation occurring first at small $b$.

A solution of  \eref{equa3} and \eref{equa4}, not the most general one, is:
\begin{equation}
\label{equa5}
\Gamma (b,R)=\frac{1}{\exp{\frac{b-R}{\gamma}+1}}.
\end{equation}

The total and elastic cross-section are written as
\begin{equation}
\label{equa6}
\sigma_{tot.}(s)=2 \pi \int \Gamma (b,R) db^2
\end{equation}
and, neglecting real part contributions,
\begin{equation}
\label{equa7}
\sigma_{el.}(s)=\pi \int | \Gamma (b,R)|^2 db^2,
\end{equation}
respectively. The imaginary part of the elastic amplitude $ {\rm Im} \, F(t,R)$ is the Fourier-Bessel transform of $\Gamma (b,R)$ and the differential elastic cross-section is written as
\begin{equation}
\label{equa8}
\frac{d \sigma}{d t}=\frac{\sigma_{tot.}^2}{16 \pi} \frac{ {\rm Im} \, F(t,R)^2}{ {\rm Im} \,F(0,R) ^2}.
\end{equation}

It should be noticed that in regions of energy where $ \gamma / R \simeq const.$, as it happens at ISR energies ($20 \lesssim \sqrt{s} \lesssim 60$~GeV), $\Gamma (b,R)$ satisfies geometrical scaling \cite{ref7},
\begin{equation}
\label{equa9}
\Gamma (b,R) \, \, \,  \raisebox{-5pt}{$\overrightarrow{\scriptstyle \gamma / R \simeq const.}$} \, \, \, \Gamma ( \beta ),
\end{equation}
with
\begin{equation}
\label{equa10}
\beta \equiv \frac{b}{R},
\end{equation}
and, 
\begin{equation}
\label{equa11}
\frac{d \sigma}{d t} \sim R^2 \left( f(tR^2) \right)^2 . 
\end{equation}

As $\gamma$, contrary to $R$, does not show in general a consistent dependence on energy, in the limit $R  \rightarrow \infty$, $\gamma / R  \rightarrow 0$ and one obtains again scaling  \cite{ref8},
\begin{equation}
\label{equa12}
\Gamma(b,R) \, \, \, \raisebox{-5pt}{$\overrightarrow{ \scriptstyle R  \rightarrow \infty }$}\, \, \, \Gamma( \beta ) \equiv 
\cases{1, & $\; \;  0 \leq  \beta < 1$\\
0, & $\; \; \phantom{0 \leq} \;  \beta > 1$} \; \; ,
\end{equation}
with $\sigma_{tot.} \sim \sigma_{el.} \sim R^2$ and $\sigma_{el.}/ \sigma_{tot.} \rightarrow const. =1 / 2 $.

It should be also noticed that the parameter $R$ in \eref{equa5} separates the region of negative curvature, $b < R$, from the region of positive curvature, $b > R$. In fact, $R$ plays the role of the angular momentum L, used in the proof of Froissart bound \cite{ref9} by Martin and collaborators \cite{ref10}, that separates the region that contributes in a significant way to the total cross-section, $\ell < L$, from the region that is negligible, $\ell > L$:
\begin{equation}
\label{equa13}
R=\frac{2}{\sqrt{s}}L=\frac{1}{\sqrt{t_0}}{\rm ln} \left( \frac{s}{s_0} \right)  ,
\end{equation}
with $\sqrt{t_0}=2 m_{\pi}$.

When comparing our model to experimental $d \sigma / d t$ one finds that $ \gamma$ takes values always of the order $\gamma \simeq 1 \; {\rm mb}^{1 / 2}$, while $R$ is a monotonically increasing function of energy. 

In fact $\gamma$, the parameter that controls the low density region, $b>R$, can be seen as a measure of the range of the interaction, with $\gamma \sim (2 m_{\pi})^{-1}$ in the Yukawa picture. The evolution of the cross-sections with the energy is controlled by the single parameter $R(s)$, the effective impact parameter radius.

When studying the evolution of the amplitude \eref{equa5} with energy one finds three regimes (see Fig. \ref{fig1}):
\begin{figure}[t]
\begin{center}
\hskip -0.9 true cm
\includegraphics[width= 9.5 cm]{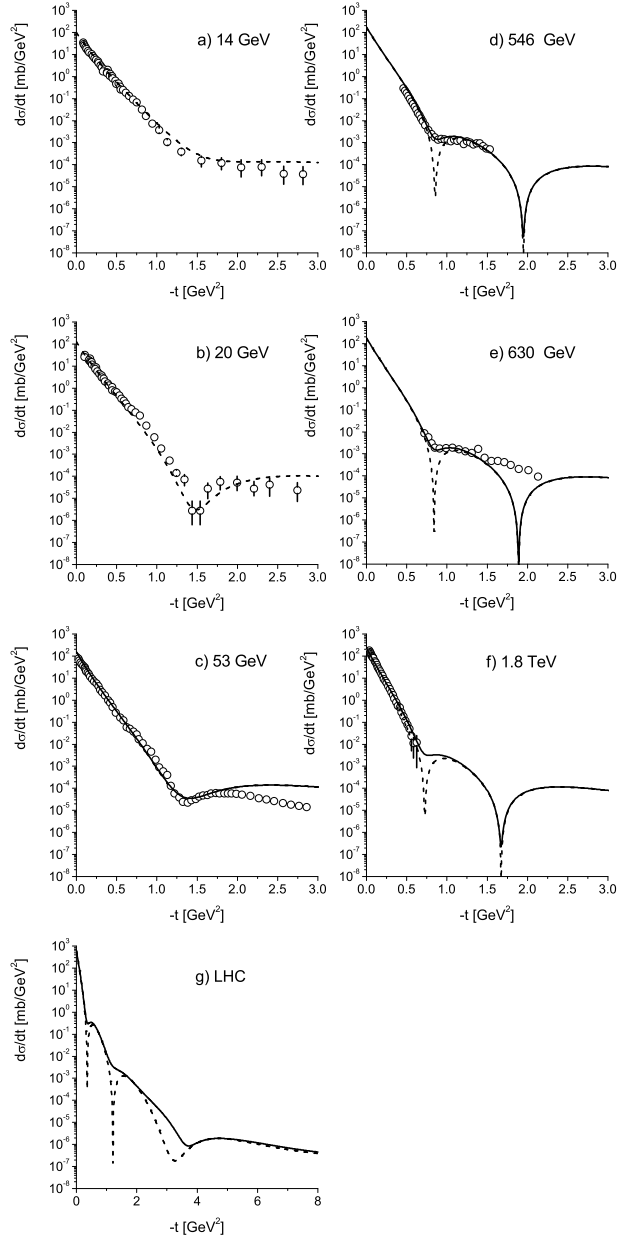}      
\end{center}
\caption{\label{fig1} $d \sigma / d t$ as function of $-t$ at different energies, showing the sequence: no structure in a), one minimum in b) and c), two zeros in d) and e) and two zeros and one minimum in f) at LHC. Values of $\gamma$ and $R$:
a) $\gamma=1.020$, $R=1.972$; 
b) $\gamma=1.026$, $R=2.171$;
c) $\gamma=1.090$, $R=2.259$;
d) $\gamma=1.000$, $R=2.575$;
e) $\gamma=1.016$, $R=2.496$;
f) $\gamma=1.078$, $R=2.683$;
g) $\gamma=1$, $R=3.770$. 
Data from \cite{ref11}. Dashed line: only imaginary part contribution. Full line: the real part of the amplitude is included.}
\end{figure}

i) $ \sqrt{s} \lesssim 20 \; {\rm GeV}$, Fig. \ref{fig1}.a).

This is the region corresponding to linear evolution, with $\Gamma$ small and with exponential behavior, $d \sigma / d t$ being a monotonically decreasing function of $-t$.

ii) $  20 \lesssim \sqrt{s} \lesssim 63 \; {\rm GeV}$ (ISR energies), Figs. \ref{fig1}.b) and c).

In this region  a dip, which is a minimum, not a zero, appears at $-t \approx 1.4 \; {\rm GeV}^2$ and slowly moves to the left as energy increases. Conventional wisdom says that the dip results from a zero: interference between one-Pomeron and two-Pomeron exchanges \cite{ref12}.

iii) $  500 \lesssim \sqrt{s} \lesssim 1.8 \; {\rm TeV}$, Figs. \ref{fig1}.d), e) and f).

In this region the minimum becomes negative, originating a pair of zeros. Instead of the clean second maximum of region ii) one has now a kind of shoulder, but with a cross-section higher by an order of magnitude.

In Fig. \ref{fig1}.g) we have also included our expectation for LHC (assuming $ \sigma_{tot.} \approx 110 \; {\rm mb}$ - see \cite{ref12a} for expected range of values - and $\gamma = 1 \; {\rm mb}^{1 / 2}$). Our LHC curve clearly shows how the evolution towards the black disk continues: from a monotonically decreasing curve at large $-t$ a minimum starts developing which at some stage generates a pair of zeros to join the previous pair. And so on! In the black disk limit we just have a sequence of pairs of zeros. 

At high energy, $ \sqrt{s} \gtrsim 60 \; {\rm GeV}$, when $\sigma_{pp} \simeq \sigma_{\bar{p} p}$ is not difficult to use the derivative dispersion relations \cite{ref12b} to estimate the real part contribution to the differential cross section. In Figs. \ref{fig1}.c) to \ref{fig1}.g) we show, in full line, $d \sigma / d t$ including the real part correction. The real part contribution is important at the zeros of $\Gamma$.

\begin{figure}[t]
\begin{center}
\hskip -0.5 true cm
\includegraphics[width= 9 cm]{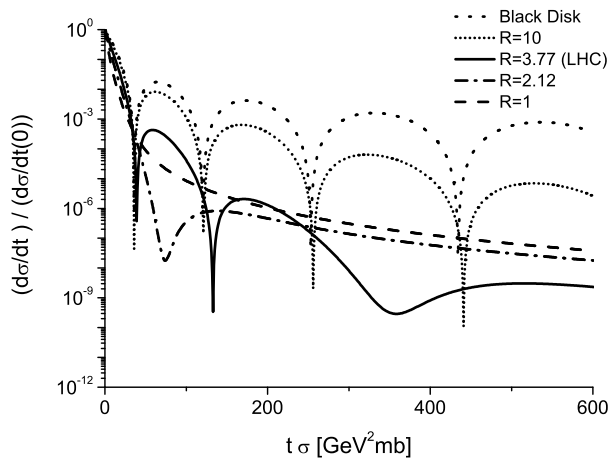}      
\end{center}
\caption{\label{fig2}$d( \sigma / d t) \; / \;( d \sigma / d t(0))$ as a function of the scaling variable $|t| \sigma_{tot.}$ showing the approach to black disk geometrical scaling from small to larger values of the scaling variable. Scaling only applies to the imaginary part of the amplitude. The parameter $\gamma$ was put equal $1$.}
\end{figure}

In Fig. \ref{fig2} we show the geometrical scaling plot (see \eref{equa11}) of $\frac{d \sigma}{d t} / \frac{d \sigma}{d t}(t=0)$ as function of the variable $| t | \sigma_{tot.} $ for different values of $R$ and for $\gamma = 1 \; {\rm mb}^{1 / 2}$, and the black disk limit. The way the approach to the scaling curve is achieved seems clear: as the energy, or $R$, increases scaling is satisfied for larger $| t | \sigma_{tot.}$ values. The LHC curve corresponds to $\sigma_{tot.}=110 \; {\rm mb}$, $\sigma_{el.}/\sigma_{tot.} \simeq 0.28$.

\begin{figure}[t]
\begin{center}
\hskip -0.5 true cm
\includegraphics[width= 9 cm]{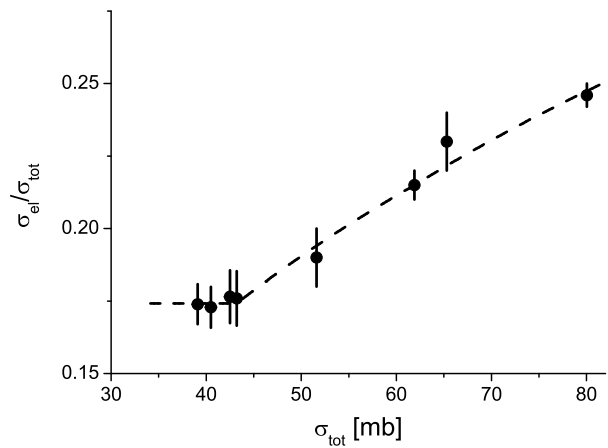}      
\end{center}
\caption{\label{fig3}$\sigma_{el.}/\sigma_{tot.}$ as a function of $\sigma_{tot.}$ making use of \eref{equa5}. Data points, from left to right: 
$\sqrt{s}=23.5 \; {\rm GeV}$, $\gamma=1.065$, $R=1.710$; 
$\sqrt{s}=30.6 \; {\rm GeV}$, $\gamma=1.090$, $R=1.731$;
$\sqrt{s}=44.9 \; {\rm GeV}$, $\gamma=1.095$, $R=1.805$;
$\sqrt{s}=52.8 \; {\rm GeV}$, $\gamma=1.107$, $R=1.810$;
$\sqrt{s}=200 \; {\rm GeV}$, $\gamma=1.123$, $R=2.100$;
$\sqrt{s}=540 \; {\rm GeV}$, $\gamma=1.080$, $R=2.498$;
$\sqrt{s}=900 \; {\rm GeV}$, $\gamma=1.026$, $R=2.662$;
$\sqrt{s}=1800 \; {\rm GeV}$, $\gamma=1.046$, $R=3.050$.
We expect for LHC $\sigma_{el.}/ \sigma_{tot.}=0.28 $ and $\sigma_{tot.}=110 \; {\rm mb}$. Data from \cite{ref11}.}
\end{figure}

In Fig. \ref{fig3} we show the correlation between $\sigma_{el.} / \sigma_{tot.}$ and $\sigma_{tot.}$ in comparison with data. 
Note the transient geometrical scaling at ISR energies, $\sigma_{el.} / \sigma_{tot.} \simeq const.$.

It should be mentioned that the curves shown in Fig. \ref{fig1} are not fits to data, but represent qualitative descriptions. This means that $\sigma_{el.}$ and $\sigma_{tot.}$, which are controlled by the first points in the low $\mid t \mid$ region, are not constrained. That is the reason why the values of $R(s)$ in Fig. \ref{fig1} do not coincide with the values used in Fig. \ref{fig3}. For instance, at ISR energies, in Fig. \ref{fig1}, $\sigma_{el.}$ and $\sigma_{tot.}$ are larger by a factor of the order of $15 \%$ relative to the true values (in Fig. \ref{fig3}).

Recently, in several papers \cite{ref13, ref14, ref15} the questions of soft physics were addressed from different points of view. However, contrary to the present paper, no attempts were made to explain the qualitative aspects of $d \sigma / d t $ evolution with energy. Instead, interesting problems related to inelastic diffraction, as Higgs production, were addressed. The black disk perspective varied significantly from \cite{ref13} to \cite{ref14}: the black disk limit being already present at LHC energies in \cite{ref13}, but occuring at extremely high energies in \cite{ref14}. For a discussion on the black disk limit see also \cite{ref15a}.

The relevance or not of the Froissart bound (see \cite{ref16} and \cite{ref17}) and a dynamical interpretation of it (see \cite{ref18} and \cite{ref19}) are still matters open to discussion.

Finally, we summarize our work. Starting from the non-linear logistic equation we obtained a solution for the high-energy imaginary part of the amplitude, and we were able to describe in a qualitative and semi-quantitative way the essential features of the evolution of the differential elastic cross-section with energy, namely the sequence: no structure in $| t | $, one minimum, two zeros and so on. Our prediction for $d \sigma / d t$ at LHC energies is different from all the ones we are aware of (see, for instance, \cite{ref20}).

\ack{
We would like to thank Carlos Pajares and Jaime Alvarez-Muniz for discussions on cosmic rays at very high energy. Special thanks are due to Teresa Pe\~ na for information on zeros and minima in nuclear reactions}

\end{document}